\documentclass[a4paper,11pt]{article}
\pdfoutput=1 

\usepackage{jinstpub} 

\usepackage{lineno}
\notoc{\notoctrue}

\title{A Custom Discrete Amplifier-Shaper-Discriminator Circuit for the Drift Chambers of the R3B Experiment at GSI }

\author{Michael Wiebusch,}
\author{Henning Heggen}
\author{and Michael Heil}
\affiliation{GSI Helmholtzzentrum für Schwerionenforschung,\\Planckstr. 1, 64291 Darmstadt, Germany}

\emailAdd{m.wiebusch@gsi.de}

\abstract{This contribution presents a pragmatic approach to read-out electronics for drift chambers used in particle physics experiments, specifically for the R3B experiment at GSI. The design uses discrete miniature SMD components and LVDS inputs of a low-cost FPGA to achieve a performance similar to classic ASIC solutions to the problem. The circuit comprises a high gain, low noise amplifier, a custom signal shaper, tailored to the specifics of proportional counter signals, and a leading-edge discriminator with programmable threshold. The presented approach offers an attractive solution for small to medium sized detector systems that require specialized read-out electronics but cannot afford the high cost and development effort associated with ASICs.}


\proceeding{
TWEPP 2023 Topical Workshop on Electronics for Particle Physics\\
Oct 1-6, 2023\\
Geremeas, Sardinia, Italy}

\begin{document}
\maketitle
\flushbottom

\section{Drift Chambers}
Drift chambers, though not a recent innovation, are prized for certain specialized applications due to their distinctive characteristics, particularly their exceptionally low material budget.
They employ a gas mixture, primarily composed of $Ar/CO_2$, as the active medium. When energetic charged particles pass through the gas, they leave behind a trail of ionized gas molecules and free electrons, the latter are then transformed into detectable electrical signals through the application of strong electric fields.
A drift chamber consists of numerous elongated drift cells within the same gas volume, usually arranged in a planar configuration.
Each cell possesses a central wire, the sense wire, surrounded by multiple field wires. Applying a positive potential on the order of $2\,kV$ between sense wire and field wires, creates an electric field which strongly increases towards the surface of the sense wire. 
The free electrons in the ionization trail experience an attractive force towards the center of the cell and drift to the sense wire as soon as the ionization occurs. If the gas mixture is chosen properly, the drift velocity of the electrons remains almost constant (at circa $70\,\mu m/ns$), even though the electric field strength increases by orders of magnitude along the radial path to the center.
Just a few micrometers away from the sense wire, the field strength becomes sufficient to initiate an ionization avalanche. Provided that the high voltage and gas mixture are suitably tuned, this avalanche is self-quenching and contained near the wire, serving as a physical charge multiplication process that amplifies the signal charge by a factor of $10^4$ and higher\cite{sauli}, while maintaining a reasonable proportionality between the primary ionization charge from the particle track and the output signal of the drift cells.
The described principle also pertains to a proportional counter. What sets a drift chamber, and its modern counterpart, the straw detector apart, is the deliberate measurement of electron drift time to determine the radial distance of a particle track from the sense wire. This is accomplished by comparing the time of arrival (TOA) of the drift chamber pulse with a fast reference detector.
By combining the information from multiple cells, particle tracking achieves a spatial resolution of few hundreds of micrometers, significantly finer than the spacing of the sense wires ($\approx 1\,cm$).
Additionally, the measurement of pulse charge provides insights into the particle's energy loss within the medium, enabling, to a certain extent, the identification of the particle species.
The signal charge equals the amount of electrons created in the avalanche and collected by the sense wire. However, these electrons do not reach the read-out electronics all at once. The cell's signal current possesses a peculiar time dependence of the form $I(t) = c \cdot \frac{1}{t_0 + t}$, with $c$, $t_0$ depending on the high voltage setting, gas properties and cell geometry. This can be understood as the effect of the positively charged gas avalanche ions drifting away from the sense wire in the cylinder capacitor-like field, continuously releasing their electrostatic grip on the signal electrons on the wire. The impulse response of a drift chamber (response to a single primary ionization charge) is a sharp pulse, few ns wide, with a fat, i.e. slower-than-exponentially decaying tail, that keeps releasing a significant fraction of the pulse charge over the course of microseconds\cite{blum_rolandi} and distorts the baseline for ensuing pulses.

\section{Analog Read-Out Electronics Requirements}

\begin{figure}
\centering
\includegraphics[width=0.7\textwidth]{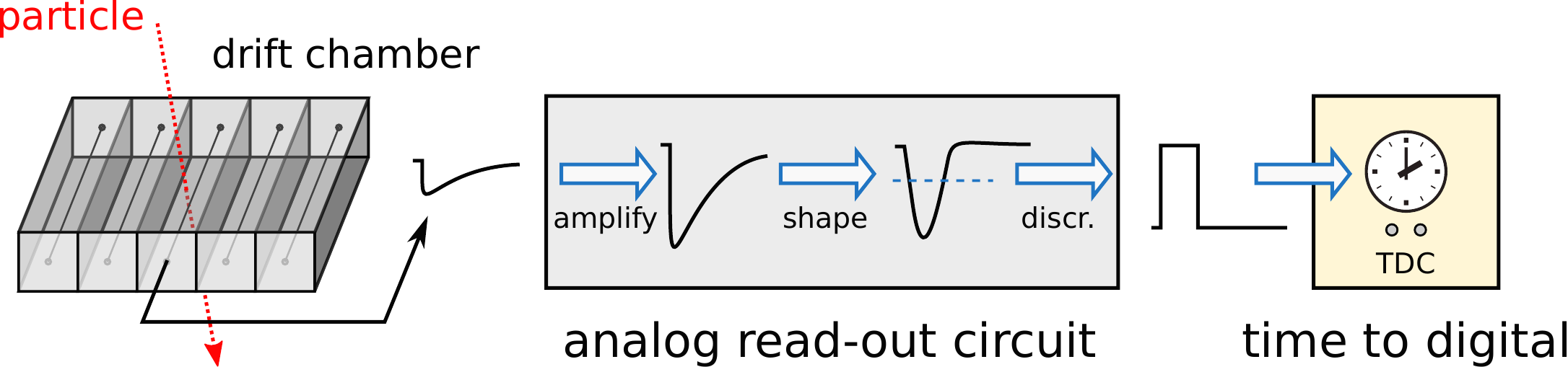}
\caption{An illustration of the Amplifier-Shaper-Discriminator signal chain.}
\label{fig:signal_chain}
\end{figure}

While drift chambers already have an internal charge amplification mechanism,
a high gain, low noise electronic amplifier is yet required to convert the sub-$pC$ charge pulses to voltage pulses with a sufficient amplitude to be processed by the subsequent signal chain.
This is a necessary first step for all types of drift chamber read-out electronics, nevertheless
there are different methods of digitizing the analog signal.
While it is possible to directly sample the entire waveform of the amplified chamber signal with an ADC, in this work the classical approach of using an Amplifier-Shaper-Discriminator + Time-to-Digital Converter (TDC) was chosen (see figure \ref{fig:signal_chain}). This concept has the inherent benefit of greatly reducing the amount of recorded data already in the front-end, since for every detector pulse only two timestamps are recorded by the TDC: The time of the leading and the trailing edge of the logic signal after the discriminator (i.e. comparator).
The leading edge contains the drift time information and yields spatial knowledge about the recorded particle track. The pulse width, i.e. the time over threshold (TOT) is a function of the pulse amplitude after the shaper, which in turn is a measure of the received charge.

For optimal performance the shaper must limit the bandwidth towards high frequencies to reduce the susceptibility to high frequency electrical interference. A second crucial function of the shaper as a low-pass filter is to act as an integrator for signal components with frequencies above the filter's corner frequency. A particle track in a drift chamber cell creates a multitude of ionization charges which arrive successively at the sense wire over several nanoseconds, each creating a short pulse which is too weak for detection. The integrating shaper allows for these pulses to add up to a single, much smoother pulse which clearly rises above the baseline noise. In practice, higher order low-pass filters are used with three or more poles. It is possible to include analog circuits known as pole-zero filters\cite{boie} in the shaper which suppress the $1/t$ shaped tail inherent to all drift chamber pulses. This ensures that the pulse waveform returns to the baseline after few hundreds of nanoseconds and thus enables the system to detect an ensuing particle track. Furthermore, shortening the tail allows for a cleaner trailing edge (transition below the threshold) and a more benevolent analytic dependence between the registered charge and the measured TOT. Ideally, the remaining tail has a roughly exponential shape, so that the time over threshold increases like the logarithm of the charge.

\section{Implementation and Performance}
\label{sec:implementation}

\begin{figure}
\centering
\includegraphics[width=1\textwidth]{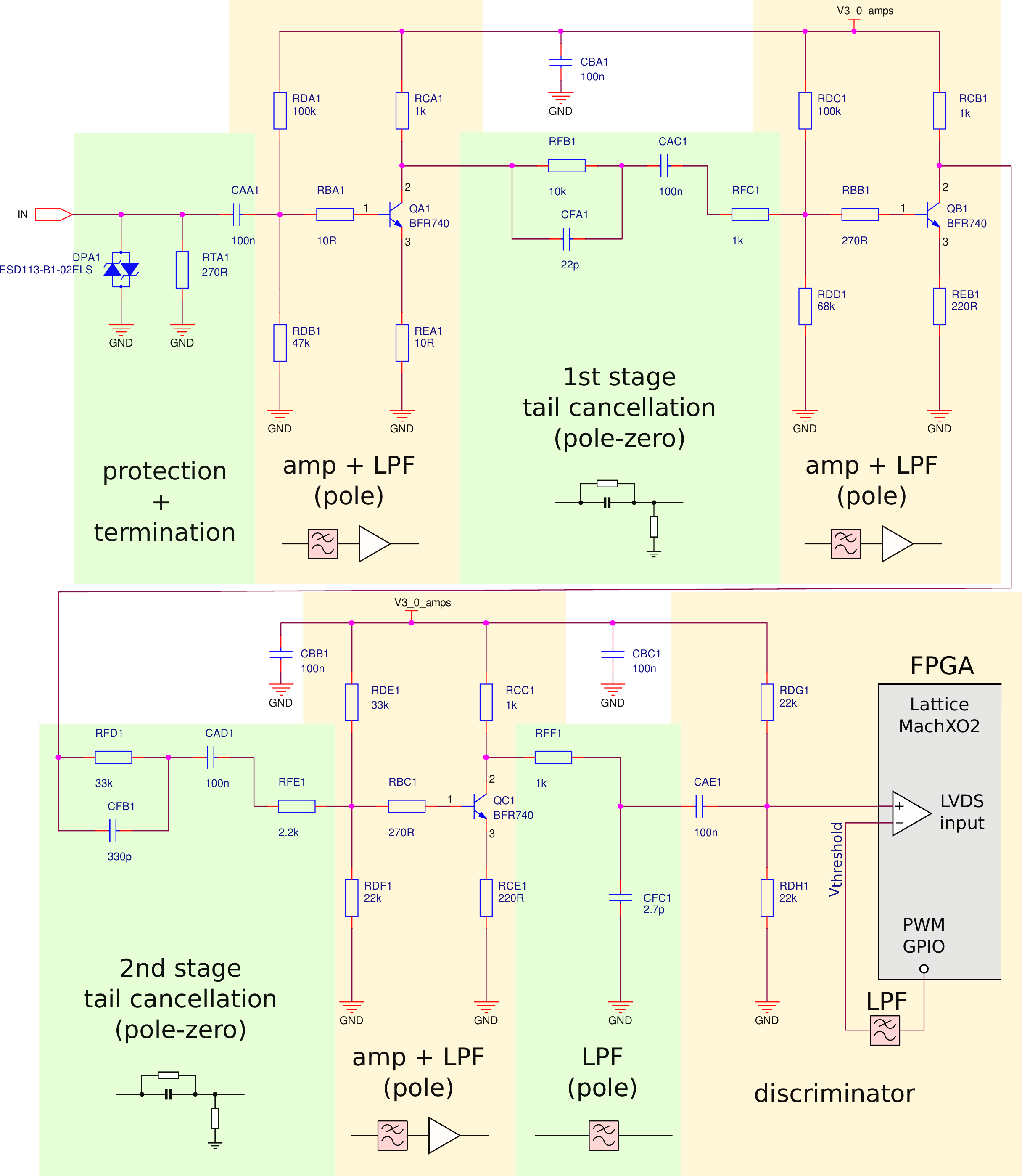}
\caption{Schematic of the Amplifier Shaper Discriminator Circuit.}
\label{fig:schematic}
\end{figure}

\begin{figure}
\centering
\includegraphics[width=0.7\textwidth]{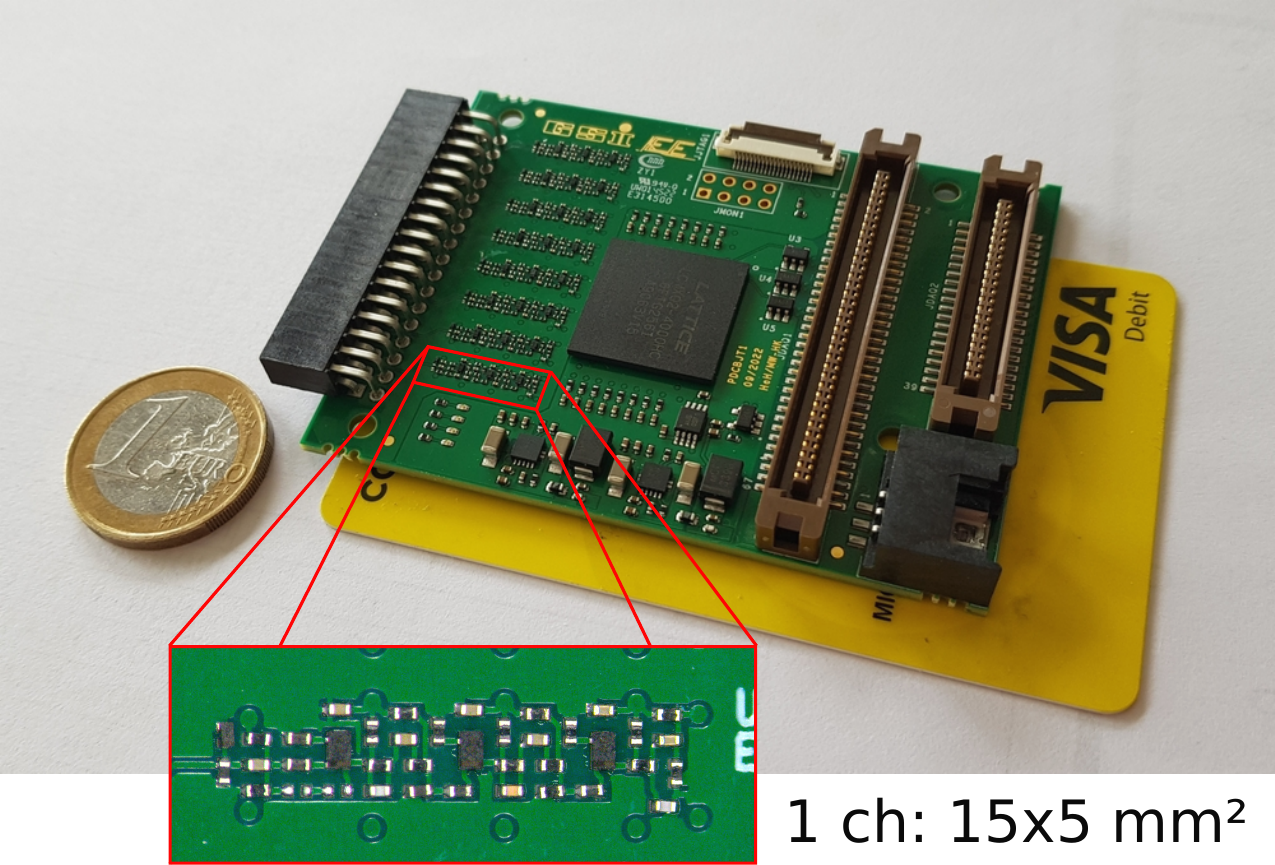}
\caption{The fully assembled printed circuit board hosting 16 amplifier-shaper circuits and an FPGA for signal discrimination.}
\label{fig:board}
\end{figure}


The aim of this work was to manufacture front-end electronics with similar performance to the classic ASD-8\cite{asd8_main} ASIC, which is no longer available due to its outdated manufacturing process, and to equip all 512 channels of the already existing drift chambers of the R3B experiment at GSI.
The design presented here is an amplifier-shaper-discriminator (ASD) circuit implemented in discrete miniature SMD technology.
Sixteen analog channels are integrated onto a single credit card-sized board (see figure~\ref{fig:board}) which connects directly to the chamber via a double row $2.54\,mm$ pin header connector. On the rear end, the board features a ribbon cable connector which carries the discriminator output pulses as twisted pair LVDS signals to a multi purpose TDC and slow control board which was developed in-house, but which is beyond the scope of this presentation.
Each amplifier-shaper channel takes up $15\times5mm^2$ of board space and comprises three individual transistors in a TSLP-3 package (size of a 0402 resistor) as well as a number of passive 0201 components.

In practice the amplifier and the shaper are inseparably intertwined (see figure~\ref{fig:schematic}).
The transistors are wired as common emitter amplifiers with emitter degeneration.
The dominant behaviour of the shaper is that of a 4th order low-pass filter.
Three of the four poles can be identified with the intrinsic frequency cut-off of the amplifier stages, while the fourth pole is achieved by an explicit RC filter.
Two passive pole-zero filters for tail suppression are placed in the intermediate places between the three amplifier stages. The high input impedance of the amplifiers prevents the filters from loading each other.
The amplifier-shaper in its entirety displays a peaking time of $14\,ns$ when stimulated with a delta-like charge pulse.
Due to its integrating behaviour the circuit acts like a charge sensitive amplifer with a gain of $3.5\,mV/fC$,
although technically being a voltage amplifier with no capacitive feedback.
The feedback-less topology and the macroscopic spacing between the gain stages has the positive side-effect of not being prone to self-oscillation, which is an often observed problem in comparable ASIC solutions.
The otherwise high input impedance of the first amplifier is reduced to $270\,\Omega$ by means of a termination resistor, chosen to match the transmission line impedance of the drift cell.
Pulse discrimination for all 16 channels is carried out with the help of the numerous low-voltage differential signaling (LVDS) inputs of a low-cost FPGA (Lattice MachXO2), which double as fast comparators when biased properly near $VCC/2$.
The threshold voltage for each comparator is generated by means of outputting a PWM waveform on a GPIO pin of the same FPGA. Low-pass filtering the PWM signal with a simple passive two-stage RC filter yields a stable DC voltage that can be tuned by slow-control with sub-$mV$ accuracy.
The total power consumption of the ASD circuit is only $36\,mW$ per channel (board power consumption/number of channels), while the transistor circuit alone accounts for $16\,mW$ per channel.
With sufficiently low threshold the circuit reliably registers pulses with a charge content $\ge 20\,fC$ safely up to $10\,pC$.
During the production of the read-out boards to equip the 512 channels of the R3B drift chambers, circa 15\% of the boards displayed faulty channels. However, it was possible to repair all channels without exception by resolving tombstone-type soldering issues with very moderate effort.
Test measurements with a pulse generator, programmed to mimic a drift chamber pulse, including the $1/t$ tail, reveal that the shaper functions according to its design, insofar as the time over threshold increases logarithmically with the input charge.
An intrinsic electronic charge measurement precision of 10\% is achieved, which is acceptable in comparison with the limited charge precision of the drift cell.
First tests with the drift chamber and new electronics using cosmic muons reveal a particle detection efficiency of $\ge 95\%$.



\section{Conclusion}
\label{sec:conclusion}
The presented approach demonstrates that modern miniature SMD components and minimalist analog circuit design can be utilized to develop analog read-out solutions that can compete with ASIC solutions, at least in moderately high density detector set-ups. 
A new custom analog front-end module was manufactured to equip all 512 channels of the R3B drift chambers.
The design requirements were met, with regards to charge sensitivity and the need to mitigate the characteristic shape of the signal tail.
With correcting minor soldering issues, a yield of 100\% working channels was achieved.
The suggested circuit design provides an appealing option for smaller experiments demanding custom read-out electronics for gas detectors, especially drift chambers and straw detectors, without incurring the substantial expenses and development efforts typically linked to ASICs.


%
%


\end{document}